\documentclass[12pt]{article}

\usepackage{graphicx,amscd,amsmath,amssymb,verbatim}
\usepackage[dvips]{hyperref}
\usepackage[TS1,OT1,T1]{fontenc}
\usepackage{subfig}
\newtheorem{theorem}{Theorem}
\newcommand{\proof}{{\bf Proof:} }

\begin{document}
\title{New Counting Codes for Distributed Video Coding}
\author{Axel Becker and Ka-Ming Leung\\Canon Information Systems and Research Australia
}
\date{\today}
\maketitle
\begin{abstract}
This paper introduces a new counting code. Its design was motivated by distributed video coding where, for decoding, error correction methods are applied to improve predictions. Those error corrections sometimes fail which results in decoded values worse than the initial prediction.
Our code exploits the fact that bit errors are relatively unlikely events: more than a few bit errors in a decoded pixel value are rare.  With a carefully designed counting code combined with a prediction those bit errors can be corrected and sometimes the original pixel value recovered. The error correction improves significantly.
Our new code not only maximizes the Hamming distance between adjacent (or "near-1") codewords but also between nearby (for example "near-2") codewords. This is why our code is significantly different from the well-known maximal counting sequences which have maximal average Hamming distance. Fortunately, the new counting code can be derived from Gray Codes for every code word length (i.e. bit depth).

\end{abstract}
\section{Introduction}
The idea behind of Distributed Video Coding was established in the 1970's by Slepian and Wolf
[\ref{slepianwolf}] and by Wyner and Ziv [\ref{wyner78}]. However, it was not before the year 2000 that a push was undertaken to actually establish working distributed video coding systems. In the last few years many impressive and thought provoking research publications appeared on the subject ( cp. [\ref{discoverbib}] for an extensive bibliography). To summarize all those findings would go beyond the scope of this paper. Currently, one of the biggest unsolved problems in distributed video coding is how to design error correction codes specifically for distributed video coding.\\
The overall principle in a distributed video compression system is the following: the decoder generates a  preliminary prediction of the frame to be decoded. In some systems, this "preliminary prediction" is generated by motion interpolation ([\ref{aaronraneetal}]), in some other systems by spatial interpolation ([\ref{morbee}], [\ref{Ascenso}]) or some hybrid interpolation schemes([\ref{tagliasacchi}]). 
Of course, those frame predictions are erroneous. In distributed video coding the prediction errors are interpreted as {\ it transmission} errors and error correction methods known from channel coding are applied to improve the quality of those preliminary predictions. Of course, error correction methods are probabilistic methods and even though an improvement is possible for the majority of the pixels it cannot be avoided that some pixel values are "mis-corrected" in practice; the corrected pixel value is worse than the predicted pixel value. This is why most distributed video compression systems employ a "reconstruction step"  which is often some form of thresholding. This reconstruction step decides whether the predicted pixel value or the error corrected pixel value is taken as final output pixel value. Resorting to the initially predicted pixel value rather than improving it is obviously undesirable which is why most authors compromise the simplicity of the encoder and choose more powerful and more complex error correction methods.\\
In contrast to this we found that it is much more advantageous to remap the pixel values to a new bit representation such that we can take additional advantage of the prediction. It is obvious that:
 \begin{enumerate}
\item a small prediction error is more likely than a large prediction error, and
\item a small number of mis-corrected bits in a pixel value is more likely than a larger number of mis-corrected bits.
\end{enumerate}
Our idea is simple but powerful; rearrange the bit representation of pixel values such that many error correction mistakes and a big prediction errors are unlikely to happen at the same time.  
This motivates a new binary code such that
\begin{enumerate}
\item  the Hamming distance between adjacent codewords (i.e. codewords representing pixel values differ by one) is high;
\item the Hamming distance between nearby codewords can be smaller than the Hamming distance between adjacent codewords, however, it is still large enough to facilitate later error correction;
\item the Hamming distance between codewords representing very different pixel values can be small.
\end{enumerate}
Of course, for a compression application it is desirable not to add redundancy during encoding ( e.g. increasing the number of bits per pixel). This constrains the new code to be a counting code.
What a counting code is will be explained later. To our knowledge this kind of code has never been studied or used before.\\
 This paper is organized as follows: in Section \ref{CountingCodes} we introduce the nomenclature and some mathematical background of counting codes. We will explain first how to generate our new codes and study some of their important properties. Of course, an example will be presented to illustrate our findings. In the next section, Section \ref{application}, we discuss how our new codes can be put into practice and what the advantage of our new code is. In the last section which is Section \ref{futurework}, we outline what we believe would be valuable future work on this topic.

\section{New Counting Codes}
\label{CountingCodes}
\subsection{Nomenclature and previous work}
\noindent Let $S(n|p)$ be a sequence of $p$ distinct binary $n$-tuples. The sequence
$S$ is called a {\it counting sequence} of length $n$ if all
${2^n}$ possible n-tuples are visited. In the sequel the binary n-tuples will be called codewords. Of course a codeword is a binary representation of a pixel value. We will index codewords in the sequence $S$ from $0$ to $p-1$ and denote the $jth$ codeword by $x_j$, $0\le j < p$; the bit positions within a codeword
$x_j$ will be counted from $1$ to $n$ going from the right to the left. The rightmost bit is assumed to be the least significant bit. \\
The average Hamming distance of a sequence is the average Hamming distance between the $p$ pairs of successive (or "near-1") codewords.  It is obvious that a cyclic Gray sequence (a sequence of codewords where each pair of successive codewords, including the pair of the first and the last codeword, differ in only one bit) has an average Hamming distance of one for example. For many applications sequences with maximum average Hamming distance are important. The following theorem establishes bounds for the Hamming distance if the sequence is a counting sequence:
\begin{theorem}
The average Hamming distance $d_{\cal H}$of a counting sequence $S(n)$, $n> 1$ is bounded according to
\begin{equation}
1\le  d_{\cal H} \le \left (n-\frac{1}{2}\right ).
\end{equation}
\end{theorem}
A proof for this theorem can be found in [\ref{vZanten}] and [\ref{Robinson}]. A counting sequence with average Hamming distance $n-\frac{1}{2}$ is called {\it maximum counting sequence}. An example for a maximal counting sequence is:
\[
{ 000,111,001,110,011,100,010,101 }.
\]
It can be seen easily that in order to achieve a maximum average Hamming distance with a counting sequence the Hamming distance of near-1 codewords must alternate between $n$ and $n-1$.
The following was proven in [\ref{Robinson}]:
\begin{theorem}
A maximum counting sequence exists for all $n>1$.\\
\end{theorem}
However, for an application in distributed video coding maximizing the near-1 Hamming distance is not enough, e.g. near-2 distances have to be considered as well.
\subsection{New Code Generation}
It took several attempts to find a code that satisfies the above design. Rather than subjecting the reader to all the mistakes we made and dead ends we headed down while developing our code we will simply present our current code,  study its properties, and illustrate with an example how it is applied later in Section \ref{application}.

In the previous section we saw that a maximum counting sequence protects mainly neighboring codewords against bit errors. Within this section we will now outline how to derive a new code of length $n$ from binary-reflected Gray codes of length $n-1$ such that near-1 neighbors are sufficiently protected as well as near-2 neighbors.  Of course, there are many ways to construct similar codes, however, we have decided to present one which is easy to derive.\\
First, we will quickly repeat how a Gray code of length $n-1$  is generated by the well known binary reflection method: we start with an initial Gray code of length 1 which is $(0,1)$. Then, the initial Gray code  is listed in reverse order which results in $(1,0)$. Next, the initial Gray code and the reverse listed code are concatenated. This results in $(0,1,1,0)$.  The length of each codeword now gets increased; the initial Gray code gets the prefix $0$ whereas the reverse listed code gets the prefix $1$. This results in the code $(00,01,11,10)$. this is the binary-reflected Gray code of length $n=2$. This process is iterated until it results in a binary-reflected Gray code of length $n-1$.  Let this code be $(x_0,x_1,x_2,\dots, x_{p-1})$.  To derive our new code this binary-reflected Gray code of length $n-1$ is listed in reverse order and concatenated to the original code again. This results in $(x_0,x_1,x_2,\dots, x_{p-1},  x_{p-1},x_{p-2}, \dots, x_1,x_0)$.  Next,  every second codeword $x_{2j+1}$ for $j=0,\dots,2^{n-1}-1$ is bitwise complemented. This results in the sequence $(x_0,{\cal C} x_1,x_2,{\cal C} x_3 \dots, {\cal C }x_{p-1},  x_{p-1},{\cal C} x_{p-2}, \dots, x_1,{\cal C} x_0)$ where ${\cal C}x$ denotes the bitwise complement of $x$.  Next, the code gets an alternating prefix throughout the sequence which results in
\[
(0x_0,1{\cal C} x_1,0x_2,1{\cal C} x_3 \dots, 1{\cal C }x_{p-1}, 0 x_{p-1},1{\cal C} x_{p-2}, \dots, 0x_1,1{\cal C} x_0).
\]
This is our new code of code length $n$.
\subsection{Example}
\label{example1}
To make our example short we assume a bit depth of $n=4$ only and show the generation of the counting code in Table \ref{generationtable}.

\begin{table}
\caption{Generation of the new counting code}
\begin{center}
\begin{tabular}{|c|c|c|c|}\hline
\label{generationtable}
{\bf Start} & {\bf Step 1} & {\bf Step 2} & {\bf Step 3} \\
Binary reflective & & & \\
Gray code (n-1=3) & Mirror & Take complements & Add prefixes \\
\hline\hline
000 & 000 & 000          &  {\bf 0}000\\
001 & 001 &{\bf 110}   &  {\bf 1}110\\
011 & 011 & 011          &  {\bf 0}011\\
010 & 010 & {\bf 101}  &  {\bf 1}101\\
110 & 110 & 110          &  {\bf 0}110 \\
111 & 111 & {\bf 000}  &  {\bf 1}000\\
101 & 101 & 101          &  {\bf 0}101\\
100 & 100 & {\bf 011}  &  {\bf 1}011\\
        & {\bf 100}       &  100   & {\bf 0}100                \\
        & { \bf 101}       & {\bf 010}       & {\bf 1}010    \\
        & { \bf 111}        &  111              &  {\bf 0}111   \\
        & {\bf 110}       & {\bf 001}        & {\bf 1}001     \\
        & {\bf 010}        &  010              & {\bf 0}010     \\
        & {\bf 011}        & {\bf 100}       & {\bf 1}100      \\
        & {\bf 001}        & 001               & {\bf 0}001      \\
        & {\bf 000}        & {\bf 111}       & {\bf 1}111      \\
\hline
\end{tabular}
\end{center}
\end{table}
\subsection{New Code Characteristics}
\begin{theorem}
Let $(x_0,x_1,x_2,\dots, x_{2p-1})$ denote our new counting code. Then
\[
d_{\cal H} (x_{(k+1)\!\!\!\!\mod 2p},x_{k\!\!\!\!\mod 2p})\ge n-1, \forall k\in \mathbb{N}.
\]
\end{theorem}
Here, $n$ denotes the codeword length of the new code whereas $p$ is the number of the codewords 
of the Gray code from which the new code is derived. \\
\proof By construction it is
\[
(x_0,x_1,\dots, x_{2p-1})=(0y_0,1{\cal C} y_1,0y_2,1{\cal C} y_3 \dots, 1{\cal C }y_{p-1}, 0 y_{p-1},1{\cal C} y_{p-2}, \dots, 0y_1,1{\cal C} y_0),
\]
where the sequence $(y_0, \dots, y_{p-1})$ is a Gray code with n-1 bits. Thus, any two adjacent codewords differ in $1$ bit because of the prefix and in at least $n-2$ bits because one of the Gray codes was complemented. $\Box$\\
This means that near-1 neighbors are protected by a sufficiently large Hamming distance. It is interesting to note that for some near-1 codewords the Hamming distance will be equal to $n$:
$d_{\cal H}(x_0,x_{p-1})=n$ and $d_{\cal H}(x_{p-1},x_p)=n.$

We are now looking into the Hamming distance of near-2 codewords:
\begin{theorem}
\[
d_{\cal H}(x_{k\! \! \! \! \!  \mod 2p},x_{(k+2) \! \! \! \! \!  \mod 2p})=\left \{ \begin{array}{r@{\quad: \quad}l}
1 & k\!\!\!\!\! \mod p = p-2 \vee k\!\!\!\!\! \mod p = p-1 \\
2 & else
\end{array} \right.
 \]
\end{theorem}
\proof Obviously,  the prefix (the first bit) of near-2 codewords is the same. Thus, only the remaining part will contribute to the Hamming distance. If $k \!\!\! \mod p \neq p-2$ or $k \!\!\! \mod p \neq p-1$ then all three suffixes of the codewords $x_{k\!\!\!\mod  2p}, x_{(k+1)\!\!\! \mod 2p}$ and $x_{(k+2)\!\!\! \mod 2p}$ come from different Gray codes. The Hamming distance thus equals 2. If
 $k\!\! \mod p = p-2$ or $k\!\! \mod p = p-1$ then either the first two or the last two remaining parts are derived from the same Gray code. The resulting Hamming distance is therefore only one. The same argument applies to  $k\!\! \mod p = 2p-2$ and $k\!\! \mod p = 2p-1$.
 $\Box$\\
 It is interesting to note that by construction the near-2 Hamming distance does not depend on the bit depth $n$.

\subsection{Continuation of Example \ref{example1} }
In Table \ref{newcountingcodetable} we show the pixel values, the new codewords representing the pixel values,  the near-1 and the near-2 Hamming distance. There, it can be seen that there are irregularities in both the near-1 and near-2 Hamming distance: for  $k=7$ and for $k=15$ the near-1 Hamming distances increase whereas the near-2 Hamming distances decrease. Of course this is because of mirroring of the underlying Gray code.
\label{codegeneration}

\begin{table}

\caption{New counting code with its near-1 and near-2 Hamming distance.}
\begin{center}
\begin{tabular}{|c|c|c|c|}\hline
\label{newcountingcodetable}
{\bf Pixel value} & {\bf Pixel value} &  near-1& near-2 \\
{\bf k} &represented in& Hamming distance& Hamming distance\\
 &new counting code &  $d_{\cal H} (x_{(k+1)\!\!\!\!\mod 2p},x_{k\!\!\!\!\mod 2p})$& $d_{\cal H} (x_{(k+2)\!\!\!\!\mod 2p},x_{k\!\!\!\!\mod 2p})$ \\
\hline\hline
0    & 0000&3&2\\
1    & 1110&3&2\\
2    & 0011&3&2\\
3    &1101&3&2\\
4    & 0110 &3&2\\
5    &1000&3&2\\
6    & 0101&3&1\\
7    &1011&4&1\\
8    & 0100 &3&2\\
9    & 1010 &3&2\\
10  & 0111&3&2\\
11  & 1001 &3&2\\
12  & 0010 &3&2\\
13  & 1100 &3&2\\
14  &  0001&3&1\\
15  & 1111&4&1\\
\hline
\end{tabular}
\end{center}
\end{table}
\subsection{Discussion}
\label{discussion}
We have seen in the example above that there are irregularities in the Hamming distances. As mentioned above our new code was designed to be used in conjunction with a prediction. This is why the irregularity in the near-1 and the near-2 Hamming distance caused by the last and the first codeword is irrelevant for applications like distributed video coding etc.; any useful prediction will not mistake $k=0$ for $k=2p-1$ and vice versa. \\
The irregularity in the middle of the code, near $k=p$ is more of a concern; of course not the increase in the near-1 Hamming distance but that the near-2 Hamming distance drops to one only for any given bit depth $n$. The immediate question is whether those irregularities can be avoided.  The following theorem is helpful in answering this question.
\begin{theorem}
There exists no counting sequence $\{x_0,x_1,\dots,x_{p-1}\}$ such that
\[
d_{\cal H}(x_{k \!\!\!\!\mod p},x_{(k+1)\!\!\!\!\mod p})=l \, \mbox{for all} \,\,\, k\in \mathbb{N},\,\, \mbox{where}\,\, l \,\,\mbox{is even}.
\]
\end{theorem}
\proof
If $d_{\cal H}(x_{k \!\!\!\!\mod p},x_{(k+1)\!\!\!\!\mod p})$ is even then the codewords $x_{k \!\!\!\!\mod p}$ and $x_{(k+1)\!\!\!\!\mod p}$ must have the same parity (e.g. even number of zeros). This means {\it all} codewords of the code must have  the same parity ( all codewords must have an even number of zeros). This means that the code can never be a counting code (the codewords with an odd number of zeros are never visited). $\Box$\\
This was first presented in [\ref{Robinson}]. It means that a counting code with an even uniform near-1 Hamming distance is impossible. However, as it follows from our construction above it is very well possible to generate a code with an almost uniform even near-1 Hamming distance for any given bit depth $n \in \mathbb{N}$. 

\section{Application}
\label{application}
Above we have mentioned that our new code was motivated by distributed video compression. Now, we will show exactly how to take advantage of it in applications such as distributed video coding for example.  In most practical video applications the pixel bit depth equals $8$ and the pixel values range from $0$ to $255$. However,  we assume that the bit depth is $n=4$ and use the code we have generated above in Example \ref{codegeneration} for the sake of brevity. The codewords are assumed to represent pixel values between $0$ and $15$.
\subsection{Example}\label{example}Let us assume that the original pixel value is $x=7$. Now, in our new code the value $7$ is represented as the codeword $x_7=(1011)$.  This pixel value is predicted with a sufficiently accurate prediction method and the obtained prediction value (in its new binary representation) is error corrected. Let us assume this prediction is $8$ which is represented by the code word $(0100)$.  Given the low dynamic range in our example (bit depth equals only 4) a prediction mismatch of less than 2 can be considered as a poor prediction, however, it is still sufficient for our purposes as we will see. The prediction value $8$ is now error corrected which results in $(1001)$ which is $11$.  It is important to note that our proposed method does {\it not} depend on any particular method of error correction and we don't want to distract with unnecessary detail. However, some readers might find it still interesting to know that we have turned blocks of pixels (and in some other experiment transform coefficients) into bit streams by bit plane scanning. We have then used a turbo code with short constraint length to do the error correction.  The decoded value $11$ differs significantly from the predicted value $8$.  This suggests that a decoding error has occurred. Now, the four codewords  with a Hamming distance of one to the codeword $(1001)$ are considered to be the most likely candidates for the final output.  These four codewords are $(0001), (1101),(1011),(1000)$ and represent the values $14, 3, 7$ and $5$ respectively. One simple approach to find a final candidate value is to take the value closest to the predicted value as final output value. In our example $7$ is the final output value.
This last step is an analogy to syndrome-coding [\ref{pradhanramchandran}].  

In previous work on distributed video coding a popular method to construct the final output value is basically thresholding: if the predicted value and the (preliminarily) decoded value differ too much then the final output value is set to be the predicted value. So, the final output value for our example would only be $11$. It is worthwhile to note that the actual reconstruction used in systems  as e.g.   [\ref{Ascenso}] or [\ref{aaronraneetal}] are more sophisticated: the reconstruction makes use of dithering and the boundaries of quantization bins for example and the final reconstruction value would be $8$ or $9$ depending on the actual implementation.  \\

\subsection{Illustration}
Example video frames which illustrate the advantage of our counting codes are shown in Figure \ref{images}. Depicted on the left hand side are video frames which were first predicted and then error corrected with a turbo code. The video frames on the right hand side were produced by employing our counting code technique. One must assume that the error correction occasionally fails on all bit planes. Clearly visible (in printouts of those video frames) are only black and white spots in the left video frames where the error correction failed on significant bit planes. For example, when predicting a flat area (e.g. the pavement in the pedestrian video) even an excellent prediction, when bit plane wise scanned, might appear as an error burst to an error correction method such that it will fail. Of course, the white and black artifacts can easily be removed by reconstruction methods similar to thresholding, too. However, our method further corrects less obvious errors which greatly contributes to the PSNR but which are not visible in printouts of the video frames. Furthermore, it is interesting to know that we have employed the (7,5) turbo code with a block length of approximately 1KB. This is known to perform very poorly but with our new technique even this simple error correction is good enough to achieve a very successful reconstruction. To us it seems that it is not important to use a sophisticated error correction method (like e.g. more sophisticated turbo or LDPC codes, eventually with error concealment) or to construct special error correction codes for a Distributed Video Coding System but to {\em transform} the input data which is not a complex operation.

\begin{figure}[h]
  \centering
  \caption{DVC compressed and reconstructed video frames; the frames on the left were produced with a system using the (7,5) turbo code, whereas the frames on the right were produced using the same system with our proposed counting code technique. }
  \label{images}
\end{figure}


\section{Final Remarks and Future Work}
\label{futurework}
In Example \ref{codegeneration} above it was mentioned that the irregularity in the middle of the code is a weakness of our code. However, in practical applications where the bit depth is higher than $n=4$ the code is longer. Thus the irregularities can easily be shifted to represent pixel values which are less important if this is necessary. \\
In Section \ref{discussion} we have discussed the near-1 Hamming distance and near-2 Hamming distance of our new code. We currently believe that for a larger bit depth $n$ there is no need to protect adjacent pixel values with a near-1 Hamming distance of $n-1$.  It might be worthwhile to find a code which sacrifices near-1 Hamming distance performance in favour of near-2 or even near-3 Hamming distance performance.

\section*{Acknowledgment}
We would like to thank CiSRA for supporting this work and the permission to disclose our findings. We wish to thank our colleagues for their support, interest, discussions and their patience with us.


\begin{thebibliography}{000}

\bibitem{aarongirod}A. Aaron, B. Girod, Compression with side information using turbo codes, \textit{Proc. IEEE Data Compression Conference, Snowbird, (2002), pp. 252-261}.
\label{aaron2002}

\bibitem{aaronranesettongirod} A. Aaron, S. Rane, E. Setton, B. Girod, Transform-domain Wyner--Ziv codec for video, \textit{SPIE Visual Communications and Image Processing Conference, San Jose (2004)}.
\label{aaronraneetal}

\bibitem{aaron03} A. Aaron, R. Zhang, B. Girod, Wyner-Ziv coding of motion videos, \textit{Proc. Asilomar Conference on Signals and Systems, Pacific Grove, California (Nov 2002), pp. 240-244}.
\label{aaron03}

\bibitem{Ascenso} J. Ascenso, C. Brites, F. Pereira, Improving frame interpolation with spatial motion smoothing for pixel domain distributed video coding, EURACIP Conference on Speech and Image Processing, Multimedia Communications and Services, July 2005.
\label{Ascenso}

\bibitem{girod} B. Girod, A. Aaron, S. Rane, D. Rebollo-Monedero, Distributed Video Coding, Proceedings of the IEEE, vol. 93, no. 1, pp. 71-83, January 2005.
\label{girod}

\bibitem{morbee}M. Morbee, J. Prades-Nebot, A. Pi\v{z}urica, W. Philips, Rate allocation algorithm for pixel-domain Distributed Video Coding without feedback channel, Proceedings of the ICASSP, Hawaii, 2007, Volume 1, pp. 521-524.
\label{morbee}

\bibitem{S. S Pradhan,  K. Ramchandran} S. S. Pradhan, K. Ramchandran, Distributed source coding using syndromes (DISCUS): design and construction, \textit{IEEE Trans. Inform. Theory 49 (3)} (2003), pp. 626--643.
\label{pradhanramchandran}

\bibitem{puriramchandran} R. Puri, K. Ramchandran, PRISM: A ''reversed'' multimedia coding paradigm, \textit{Proc. International Conference on Image Processing, Barcelona, Spain (Sept. 2003), Volume 1, pp. 617-620.}
\label{puriramchandran}

\bibitem{Robinson}J. P. Robinson, M. Cohn, Counting sequences, IEEE Transaction on Computers, Vol. C-30, No. 1, pp 17-23, 1981.
\label{Robinson}

\bibitem{savage}C. Savage, A survey of combinatorial Gray codes, SIAM Review, Vol. 39, Issue 4, Dec. 1997, pp. 605-629.
\label{Savage}

\bibitem{SlepianWolf} D. Slepian, J. K. Wolf,
Noiseless coding of correlated information sources, \textit{IEEE Transaction on Information Theory 19 (4)} (1973), pp 471--480. \label{slepianwolf}

\bibitem{fernando} M. Tagliasacchi, A, Trapanese, S. Tubaro, J. Ascenso, C, Brites, F. Pereira, Exploiting Spatial Redundancy in Pixel Domain Wyner-Ziv Video Coding, IEEE International Conference on Image Processing, Atlanta, October 2006, pp. 253-256.
\label{tagliasacchi}

\bibitem{vZanten} N. Suparta, A. Jan van Zanten, Balanced Maximum Counting Sequences, \textit{IEEE Transaction on Information Theorey, Vol. 52, No. 8,  August 2006, pp. 3827-3830}
\label{vZanten}

\bibitem{wyner} A. D. Wyner, The rate-distortion function for source coding with side information at the decoder II: general sources, \textit{Inf. Control 38 (1)} (1978), pp. 60-80.
\label{wyner78}

\bibitem{discoverbib} http://www.discoverdvc.org/
\label{discoverbib}
\end{thebibliography}
\end{document}